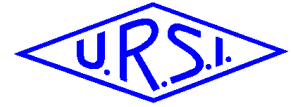

# Can Type III Radio Storms be a Source of Seed Particles to Shock Acceleration?

Nat Gopalswamy*(1), Sachiko Akiyama(2), Pertti Mäkelä(2), Seiji Yashiro(2), and Hong Xie(2)
(1) NASA Goddard Space Flight Center, Greenbelt, MD 20771, USA, https://cdaw.gsfc.nasa.gov
(2) The Catholic University of America, Washington DC 20064, USA

## Abstract

An intense type III radio storm has been disrupted by a fast halo coronal mass ejection (CME) on 2000 April 4. The CME is also associated with a large solar energetic particle (SEP) event. The storm recovers after ~10 hrs. We identified another CME that occurs on 2003 November 11 with similar CME properties but there is no type III storm in progress. The 2003 November 11 CME is also not associated with an SEP event above the background (< 2 pfu), whereas the one with type III storm has an intense SEP event (~56 pfu). One of the factors affecting the intensity of SEP events is the presence of seed particles that are accelerated by CME-driven shocks. We suggest that the type III storm source, which accelerates electrons to produce the storm, also accelerates ions that serve as seed particles to the CME shock.

## 1 Introduction

It has been well established that both source and environmental factors affect the properties of solar energetic particle (SEP) events. One of these factors is the presence of suprathermal particles in the ambient medium that can be accessed by shocks driven by particle accelerators such as a coronal mass ejections (CMEs). A recent study has demonstrated the importance of seed particles: the lower fluence level of cycle-24 SEP events is related to the lower number density of suprathermal ions in the cycle. The number density of suprathermals declined by a factor of 3 for protons and oxygen ions, and by a factor of ~7 for Fe ions when compared to the corresponding values in cycle 23 [1]. SEP events have been reported to have higher intensities when seed particles are observed at 1 AU [2-3]. They concluded that the 1-AU suprathermals are representative of seed particles in the corona where shocks accelerate particles more efficiently [4]. Moreover, the suprathermal proton population in the energy range 0.16-0.32 MeV is a better characteristic of the shock seed population than the ones in the range 1.28-2.56 MeV [3].

While considerable progress has been made in understanding the origin and properties of seed particles associated with flares and CMEs [5], another potential source of particles that has not been considered is the type III radio storms. These bursts are produced by nonthermal electrons accelerated in solar active regions bordering coronal holes. The magnetic configuration of a type III storm source is referred to as OFAR (open field lines bordering and active region) [6]. The open field lines and adjacent closed field lines should be such that they reconnect (interchange reconnection). Since most of the electron accelerators are also proton accelerators, it is logical to expect protons to be accelerated in the type III storm source. In this paper, we compare two fast and wide CMEs with very similar properties that differ in one aspect: presence of a type III storm in the active region where the CME originates. The CME associated with the type III storm is also associated with a large SEP event.

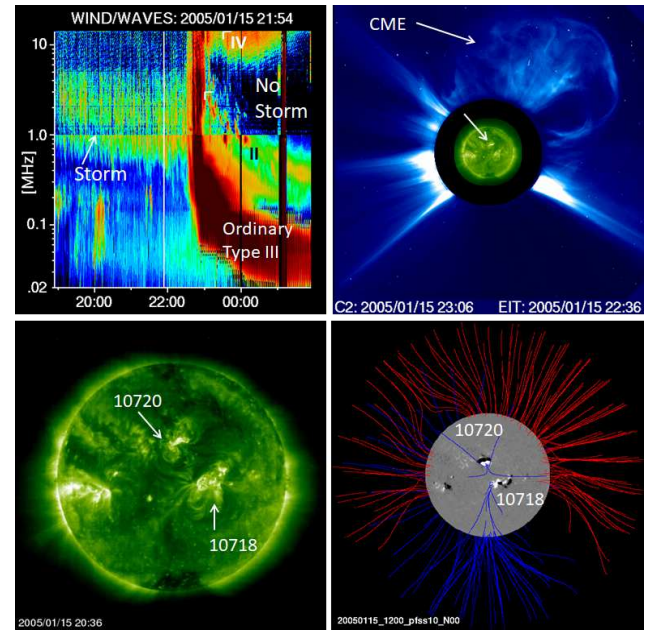

Figure 1. A type III storm observed by Wind/WAVES [7] (top left) disrupted by a large-scale eruption involving a halo CME observed by SOHO/LASCO [8] and an intense flare observed by SOHO/EIT [9] (top right). SOHO/EIT 195 Å image showing the source active region (AR 10720) that hosts the storm (bottom left). The open field lines from the potential field source surface (PFSS) extrapolation of the photospheric fields observed by SOHO/MDI [10] (bottom right); red: negative, and blue: positive field lines.

In this paper we consider two SEP events, one accompanied by a type III storm, while the other is storm free to see if the particle events differ significantly. The properties of CMEs, type III and Type II radio bursts underlying the two SEP events are similar, differing only in the intensity of the SEP events.

## 2 Type III Storm – CME Interaction

In order for the type III storm suprathermals to serve as seed particles, they need to be accessed by a CME-driven shock. That shocks do access the storm particles is evidenced by the fact that a storm can be disrupted by a CME from the same active region that hosts the storm [11-13]. It typically takes ~10 hr for the storm to reappear. Figure 3 shows a disruption event that occurred on 2005 January 15 around 23 UT [13]. The storm is hosted by NOAA AR 10720 and gets disrupted repeatedly by five CMEs between January 15 and 20. After the January 20 CME, the storm does not return. AR 10720 has the OFAR configuration as indicated by the open field lines bordering the AR. Note that the type III storm is very intense and disappears at the start of the eruption indicated by the long duration type III, type II, and type IV bursts associated with the eruption. The storm disappears immediately after the type II burst. The storm disruption demonstrates that the CME shock originating from the AR has physically passed through the storm source, thus gaining access to the suprathermal particles emanating from the AR periphery.

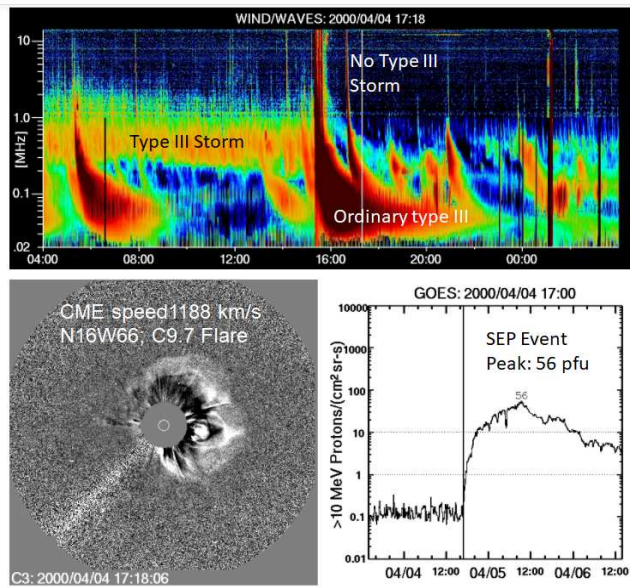

**Figure 2.** The 2000 April 4 type III storm (top) disrupted by a fast halo CME (bottom left). The fast (1188 km/s) halo CME is heading northwest from the source region at N16W66 associated with a C9.7 flare. The associated SEP event with a proton intensity of 56 pfu (bottom right) in the GOES >10 MeV channel. Storm and no-storm intervals are marked. A type II burst and an ordinary type III burst are associated with the eruption. The type III storm starts reappearing around 02:00 UT the next day, about 10 hrs after the eruption.

## 3 Type III Storms and SEP Events: A Case Study

We compare a CME associated with a large SEP event and a type III storm with a storm-free CME of similar properties. Figure 2 shows an eruption on 2000 April 4 that disrupted an ongoing type III storm observed in the Wind/WAVES dynamic spectrum. The storm recovered after about 10 hrs. The associated CME is a full halo originating from AR 8933 near the west limb (N16W66). PFSS field lines drawn on the MDI magnetogram indicate copious open field lines in the AR vicinity (Fig. 3). We think the storm originates from this region because the CME disrupts it. The fast (1188 km/s) CME is associated with a large SEP event (>10 MeV proton intensity ≥10 pfu) and a C9.7 flare. This event is selected because it is well isolated (no significant CME within preceding 24 hours from this region) and hence CME interaction is not a factor [4]. The >10 MeV SEP intensity peaks at 56 pfu and lasts for ~4 days. The CME is slightly slower than typical SEP-associated CMEs (~1500 km/s) [14].

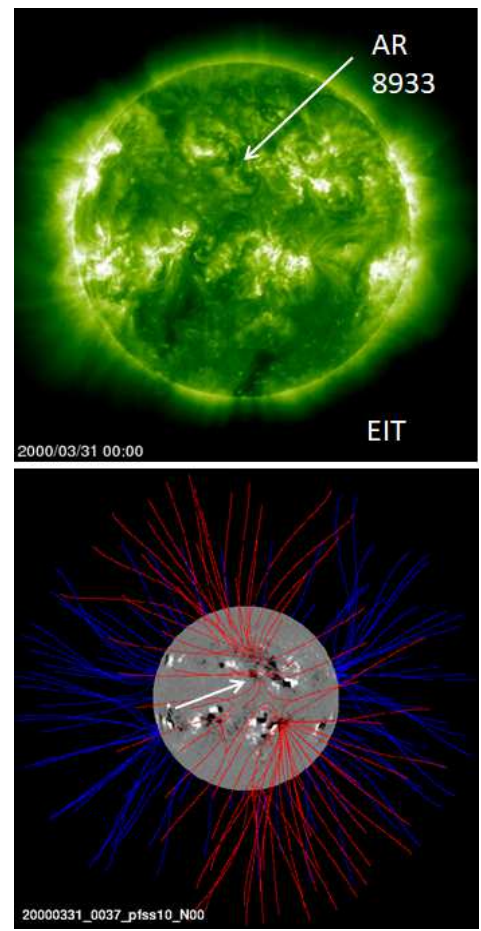

**Figure 3.** The source region of the 2000 April 4 event shown in a SOHO/EIT image taken ~4 days before the eruption when the AR is near the disk center (top). Open field lines from the northern and southern hemisphere (bottom). Since the CME disrupts the storm, we infer that the type III storm comes from the same region as the CME.

Searching for a CME with properties similar to those of the 2000 April 4 CME, we identify the fast (1315 km/s) halo on 2003 November 11 shown in Fig. 4. The CME originates from AR 10498 (S03W61) in the western hemisphere (Fig. 5) in association with an M1.6 flare. The Wind/WAVES dynamic spectrum indicates that the CME is associated with eruption-related type III and type II bursts. However, there is no type III storm in progress. The GOES proton intensity plot shows that there is no large SEP event. Thus, the primary differences between the two events are: (i) the 2003 November 11 CME is not associated with a type III storm, and (ii) there is no >10 MeV enhancement above the preexisting background intensity of ~2 pfu. Thus, the 2000 April 4 CME that disrupts the noise storm has at least an order of magnitude higher proton intensity than the storm-free CME. One way of interpreting this observation is that the 2000 April 4 CME has seed particles accelerated in the type III storm source.

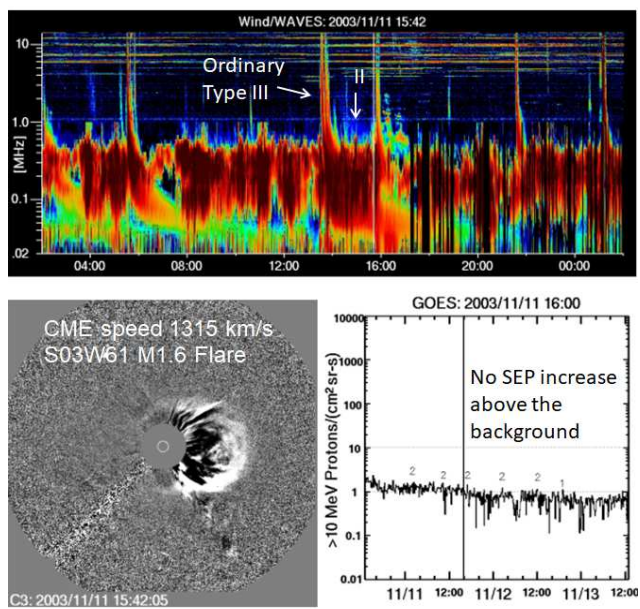

**Figure 4.** Wind/WAVES dynamic spectrum showing an ordinary type III burst and a type II burst associated with the 2003 November 11 eruption (top). There are other type III bursts in the interval shown, but no type III storm. SOHO/LASCO difference image showing the fast halo CME heading west (bottom left). The speed of the CME, its heliographic location, and the flare size are noted on the image. GOES proton intensity in the >10 MeV channel showing a pre-existing background at ~2 MeV (bottom right). The vertical line denotes the time of the CME indicating no SEP enhancement associated with the CME.

## 4 Discussion

The primary finding of this work is that a CME originating from an active region hosting a type III storm is associated with an intense SEP event, while a similar CME not associated with a type III storm has no such SEP association. The high SEP intensity from the CME that disrupted the type III storm is interpreted as a consequence of the seed particles available above the noise storm region. We appreciate that there are many factors that affect the intensity of large SEP events: the local magnetosonic speed, CME kinematics, CME interaction, presence of seed particles, magnetic connectivity, among others. Of these, we considered only the seed particle issue. Magnetic connectivity is not a factor in these events because both are from the western hemisphere of the Sun well-connected to an Earth observer. There are no preceding CMEs that may indicate CME interaction. A halo CME at 2:30 UT precedes the 2003 November 11 storm-free CME, but this CME is behind the west limb with no SEP signatures observed at Earth above the prevailing background of ~2 pfu. As for CME kinematics, the average speeds are similar, but the 2003 November 11 CME decelerates (–37 m s$^{-2}$) whereas the 2000 April 4 CME accelerates (+12.6 m s$^{-2}$) within the coronagraph FOV. However, the speeds of the two CMEs are roughly the same by the time they left the coronagraph FOV: 1232 km/s (2000 April 4) and 1090 km/s (2003 November 11). The initial acceleration computed from the flare rise time and average CME speed is higher for the 2003 November 11 CME (0.73 km s$^{-2}$) vs. 0.37 km s$^{-2}$). Assessing the magnetosonic Mach numbers above the active region corona is difficult because we do not have magnetic field measurements.

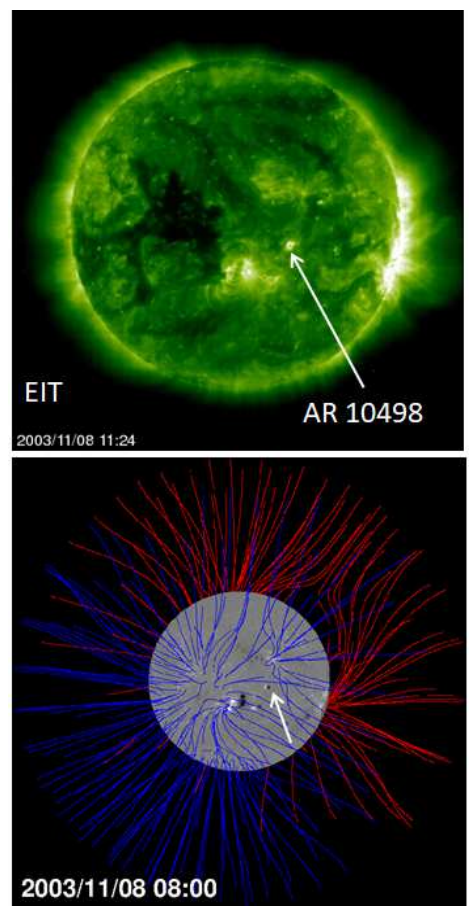

Figure 5. SOHO/EIT image showing AR 10498, which is the source of the 2003 November 11 CME (top). PFSS open field lines indicate no open field lines in the vicinity of the AR, consistent with the lack of type III storm.

The case study presented in this paper needs to be extended to more SEP events with and without an accompanying type III storm. A statistical study comparing intensities of the two sets of events should be able to provide further support to our idea that type III storm source regions are a source of seed particles to CME-driven shocks. The possibility of seed particles from type III storms supplied to the accelerator in corotating interaction regions has been considered in the accompanying paper [6].

# 5 Acknowledgements

We benefited from the open data policy of *SOHO*, *STEREO*, *SDO*, *GOES*, and *Wind* teams. Work supported by NASA's LWS program.